\documentclass[%
 reprint,
showpacs,%preprintnumbers,
 amsmath,amssymb,
 aps,
pre,
floatfix,
]{revtex4-1}

\usepackage{graphicx}% Include figure files
\usepackage{dcolumn}% Align table columns on decimal point
\usepackage{bm}% bold math
\usepackage[caption=false]{subfig}
\usepackage{algorithmicx}
\usepackage{algpseudocode}
\begin{document}

%\preprint{}

\title{Predictability of Volatility Homogenised Financial Time Series}% Force line breaks with \\

\author{Pawe\l{} Fiedor}
\email{Pawel.F.Fiedor@ieee.org}
\affiliation{%
 Cracow University of Economics\\
 Rakowicka 27, 31-510 Krak\'{o}w, Poland
}%
\author{Odd Magnus Trondrud}
\email{trondrud@stud.ntnu.no}
\affiliation{%
 Norwegian University of Science and Technology\\
 Sem S{\ae}lands vei 7-9, 7491 Trondheim, Norway
}%

\date{\today}

\begin{abstract}
Modelling financial time series as a time change of a simpler process has been proposed in various forms over the years. One of such recent approaches is called volatility homogenisation decomposition, and has been designed specifically to aid the forecasting of price changes on financial markets. The authors of this method have attempted to prove the its usefulness by applying a specific forecasting procedure and determining the effectiveness of this procedure on the decomposed time series, as compared with the original time series. This is problematic in at least two ways. First, the choice of the forecasting procedure obviously has an effect on the results, rendering them non-exhaustive. Second, the results obtained were not completely convincing, with some values falling under 50\% guessing rate. Additionally, only nine Australian stocks were being investigated, which further limits the scope of this proof. In this study we propose to find the usefulness of volatility homogenisation by calculating the predictability of the decomposed time series and comparing it to the predictability of the original time series. We are applying information-theoretic notion of entropy rate to quantify predictability, which guarantees the result is not tied to a specific method of prediction, and additionally we base our calculations on a large number of stocks from the Warsaw Stock Exchange.
\end{abstract}

\maketitle

%\tableofcontents

\section{Introduction}

Volatility homogenisation decomposition is a new method designed to aid the forecasting of changes on financial markets \cite{Kowalewski:2014}. This method explores the idea of modelling financial time series at regular points in space (i.e. price) rather than regular points in time. The authors propose that more predictive power can be extracted from the time series in such a way, as compared to standard time series describing log returns in regular time windows. It is this concept of modelling time series at regular points in space that they call volatility homogenisation. The method depends on the correct choice of quantum in terms of regular steps in space, for which there doesn't appear to be any a priori rule. Nonetheless, it's easy enough to calculate the results for various quanta and find the best suited one for a given application. The authors believe that such decomposition can replace noise interfering with prediction methods, and help to uncover the underlying patterns in the time series. Furthermore, they see this technique as a way of decoupling spatial and temporal dependence, further replacing the unnecessary noise.

The characteristics of this method have been partially proved in practical application \cite{Kowalewski:2014}. Their results show that on average volatility homogenisation decomposition yields better predictions than the same applications based on the original data (with regular points in time), particularly when applied to techniques such as Support Vector Regression (SVR) and Autoregressive Integrated Moving Averages (ARIMA) models. This is problematic in at least two ways. First, the choice of forecasting procedure (in their case SVR and ARIMA) obviously has an effect on the results, thus the efficacy of the method is proved within a very limited scope. Second, the results obtained were not completely convincing, with some predictions having under 50\% guessing rate for the decomposed data. Additionally, only nine stocks from Australian market were being investigated, which begs the question of wider application and robustness of this method. To partially remove questions arising from these omissions in the original research, we apply information theory to determine the robustness of this decomposition. We find a priori predictability of resulting times series, which does not depend on the particular method of prediction and rather speaks to the existence of patterns in the data, which could then be used by particular trading algorithms. It is then directly measuring the existence of the underlying patterns the authors of this methods mention in their argument. We estimate predictability of a process through finding inherent structural complexity of this time series, about which we do not make any assumptions. It can be a result of dimensionality, nonlinearity, non-stationarity of the generating process, or from measurement errors and finite data length. It is postulated that predictability and structural complexity are directly correlated \cite{Garland:2014}.

As mentioned, we use information theory, particularly a concept related to the Shannon entropy, its growth rate with respect to word length, or the Shannon entropy rate, to measure the complexity and unpredictability in financial time series. Time-series consisting of i.i.d. random variables (equivalently price formation processes under Efficient Market Hypothesis) have maximal entropy rates, while highly structured time-series have low entropy rates. A time series characterised by high entropy rate is highly unpredictable, and one with low entropy rate is usually reasonably predictable. Pesin's relation \cite{Pesin:1977} states that in a chaotic dynamical system, the Kolmogorov-Sinai entropy is equal to the sum of all positive Lyapunov exponents $\lambda_i$. These exponents quantify the rate at which nearby states of the system diverge with time: $\left| \Delta x(t) \right| \approx e^{\lambda t} \left| \Delta x(0) \right|$. Higher entropy is equivalent to faster divergence. Studying financial markets we are not able to estimate the Kolmogorov-Sinai entropy through Lyapunov exponents. Even ignoring this process being non-trivial in principle, most financial processes are characterised by infinite Laypunov exponents, thus estimating them is pointless. The Kolmogorov-Sinai entropy can be defined as the supremum of the Shannon entropy rates of all partitions however \cite{Petersen:1989}. Measuring spatiotemporal complexity through Shannon entropy rate requires categorical data: $x_i \in \mathcal{S}$ for some finite or countably infinite \emph{alphabet} $\mathcal{S}$. Data taken from real-world financial markets are effectively real-valued however. Thus we need to discretise the data, this is usually done by binning. This can be avoided using permutation entropy \cite{Garland:2014}, but we will use binning approach, which we postulate to be more natural and easier to interpret for financial data.

Having a method for estimating predictability through structural complexity we can then compare the predictability of the decomposed time series with the original ones without assuming a particular prediction method renders the results more general, and makes it possible for us to shed light on the usefulness of this method not only in finance, but also in any field where similar decompositions may be of use. While the method is designed to be applied to tick-by-tick data we have also applied it to the daily stock returns to see how it works with financial data at a different time scale. We investigate all stocks listed on Warsaw Stock Exchange due to large database being available to us. Additionally, in earlier studies we have shown that the Shannon's entropy rates behave similarly on this market and other important markets \cite{Fiedor:2013,Fiedor:2014}, particularly the New York Stock Exchange. In fact we also replicated this study on shorter time series for NYSE100 stocks, and the results were largely similar to the ones obtained for Warsaw's market, only less stable due to the length of the time series studied, thus we only report the results for Warsaw Stock Exchange. The results obtained in this study, having a large number of stocks investigated over long periods, should therefore be representative.

The structure of this paper is as follows. In Sect.~2 we present the method of decomposing financial time series known as volatility homogenisation and the information-theoretic tools used for assessing predictability. In Sect.~3 we describe the time series used in the empirical part, as well as describe the process and obtained results. In Sect.~4 we discuss the obtained results. Sect.~5 concludes the study and proposes further research.

\section{Methods}

In this section we first describe the volatility homogenisation decomposition, and then describe predictability estimators based on information theory. Under the Efficient Market Hypothesis, that is in a fully efficient market, there is no possibility of arbitrage. In other words Efficient Market Hypothesis presumes every market participant has all the information at every point in time. This assumption is sometimes slightly relaxed in various ways, as it is obviously at odds with the empirical data. Notwithstanding, this can be formulated in many ways. For example, it can be said that the price formation processes maximise entropy production. Alternatively, we may say that prices behave as semimartingales \cite{Delbaen:1998}. A semimartingale is a stochastic process which can be decomposed as a sum of a local martingale (representing the noise) and a finite variation term (in this case representing the drift). Consequently removing the drift leaves a continuous local martingale, which can be represented as a continuous time change of Brownian motion \cite{Dambis:1965,Dubins:1965}. That is:
\begin{equation}
\mathrm{d} X(t)=\sigma(t)\mathrm{d} W(t),
\end{equation}
where $W(t)$ is a Brownian motion and $\sigma^2(t)$ is the volatility at time $t$, which can be expressed equivalently, with natural \emph{time scale} of integrated volatility, as
\begin{equation}
X(t) = W(\theta(t))\;\textrm{where}\; \theta(t) = \int_{0}^{t} \sigma^2(s) \mathrm{d}s.
\end{equation}

Kowalewski et al. \cite{Kowalewski:2014} suppose that the market is not fully efficient, but nearly so. Incidentally, we have proven that recently for daily log returns, while intraday log returns seem to be, on average, slightly less efficient due to the clustering of volatility \cite{Fiedor:2014}, but in many cases reasonably close. In accordance with this, they propose that prices may be represented as a sum of a finite variation drift and a term close to a time-changed Brownian motion. They then concentrate on finding structure in the noise term, since the other term is straightforward to predict. They assume no jumps, getting a process of
\begin{equation}
X(t)=V(\theta(t)),
\end{equation}
where $\theta$ is a continuous time-change and $V$ is close to Brownian motion. The fundamental assumption behind volatility homogenisation is that $V$ and $\theta$ are easier objects to forecast than $X$. Thus to forecast $X$ they are separating $V$ and $\theta$ and forecast them individually. They represent $V$ as the spatial component of the process and $\theta$ as the temporal component. Incidentally, the idea of modelling a financial time series as a time-change of a simpler process is not new \cite{Mandelbrot:1967,Madan:1990,Heyde:1999}.

To decompose $X$ into $V$ and $\theta$ a spatial skeleton of the process may be used. That is, fixing some $\delta > 0$, and then considering up or down movements of size $\delta$. More precisely, if we let $T_i$ be the $i$-th time the process $X$ moves a distance $\delta$, then
\begin{equation}
T_i := \inf\{t>T_i:|X(t)-X(T_{i-1})| = \delta\},\; T_0 = 0.
\end{equation}
Also, let $\tau_i$ be the $i$-th time $V$ moves a distance $\delta$, then $X(T_i) = V (\theta(T_i)) = V (\tau_i)$, that is $\theta(T_i) = \tau_i$. An algorithm of the steps required in calculating these parameters, including the linear interpolation is presented below \cite{Kowalewski:2014}: 

\begin{algorithmic}
\Require $n \geq{} 2, X(t) = (X(t_0),X(t1),\ldots,X(t_n)),$
\State $t = (t_0, t_1, \ldots, t_n)$\\
\State $T_0 \gets t_0$
\State $X(T_0) \gets X(t_0)$
\State $i \gets 0$
\For{$j=1$ to $n$}
\If{$|X(t_j) - X(T_i)| \geq \delta$} %then\\
\If{$X(t_j) > X(T_i)$} %then
\State $X(T_{i+1}) \gets X(T_i) + \delta$
\ElsIf{$X(t_j) < X(T_i)$} %then
\State $X(T_{i+1}) \gets X(T_i) - \delta$
\EndIf
\State $T_{i+1} \gets t_{j-1} + \frac{X(T_{i+1})-X(t_{j-1})}{X(t_j)-X(t_{j-1})} \times (t_j - t_{j-1})$
\State $i \gets i+1$
\EndIf
\EndFor
\end{algorithmic}

$T_i$ can be used to get an estimate of $V$ and $\theta$. As $\delta$ approaches zero, this technique recovers the original time series. Conversely, as $\delta$ is being increased, the spatial skeleton will have any noise inherent in the original time series reduced. If $V$ were Brownian motion, then the $\tau$ would be independent and identically distributed (i.i.d.). Accordingly $\tau_i$ can be estimated by $i$ (any constant scaling of the $\tau$ would be equivalent), which gives $\hat{\theta}(T_i) = i$ as an estimate of $\theta$, and $\hat{V}(i) = X(T_i)$ as an estimate of $V$.

We know turn to the description of the estimation method of the predictability of the original and decomposed time series. In this paper, instead of trying to predict the price changes, which has the disadvantage of the results depending on the specific prognostic procedure applied, we find the predictability of the sequences themselves. This removes from our analysis the possibility that the chosen prediction method doesn't work very well with the particular kind of patterns we observe. We can do this using the Shannon's concept of uncertainty (usually called Shannon's entropy). In particular entropy rate of a dynamic process, as defined by Shannon, measures the uncertainty that remains in the next information produced by the process given complete knowledge of the past. In our context it gives the uncertainty of the next price change given complete knowledge of the past price changes. Is is therefore a natural measure of the difficulty faced in predicting the evolution of the process. Here we will define entropy and entropy rate in the Shannon's sense, as well as their estimators, which will be used in this study.

Using the Shannon's formulation of information entropy we may say that low entropy indicates high certainty and predictability in the information process, and high entropy indicates low predictability and information availability in the studied process. The predictability of a time series can naturally be estimated using Shannon's concept of entropy rate, which is a term derivative to the notion of entropy. The Shannon's entropy of a single random variable $X$ is defined as 
\begin{equation}
	\label{eq:Def_entropy}
H(X) = -\sum_{i} p(x_i) \log_2 p(x_i) 
\end{equation}
summed over all possible outcomes $\{x_i\}$ with their respective probabilities $p(x_i)$ \cite{Shannon:1948}.

Shannon also introduced the entropy rate, which generalises the notion of entropy for sequences of dependent random variables. For any stationary stochastic process $X = \{X_i\}$, the entropy rate is defined as
\begin{equation}
	\label{eq:Def_entropy_rate}
	H(X) = \lim_{n \rightarrow \infty} \frac{1}{n} H(X_1, X_2, \dots, X_n).
\end{equation}
As noted above, the right side of \eqref{eq:Def_entropy_rate} can be interpreted so that entropy rate measures the uncertainty in a quantity at time $n$ having observed the complete history up to that point. It can therefore be understood as a natural estimator of predictability.

Estimation of entropy and entropy rates is very important in many applications, as real entropy is known in just a few isolated cases. Consequently, entropy estimation has been an active field of study in the recent past, particularly with regards to the recent advances in neurobiology \cite{Maciejewski:2008}. Methods of estimation of entropy can be grouped into two distinct categories \cite{Gao:2006}: maximum likelihood or plug-in estimators and estimators based on data compression algorithms. The plug-in estimators are not suitable for analysing the mid- and long-term relations (and these are important in economics and finance), hence we are using estimators from the second group. Most of these are based on either Lempel-Ziv algorithm \cite{Farah:1995,Kontoyiannis:1998a,Lempel:1977} or Context Tree Weighting \cite{Willems:1995,Kennel:2005}. Both methods are precise even for a limited sample size \cite{Louchard:1997,Leonardi:2010,Fiedor:2014}, therefore are better equipped to deal with mid- and long-term relationships in the analysed data. In this study we use the Context Tree Weighting due to its speed.

Here we introduce the method of entropy rate estimation based on Context Tree Weighting algorithm. For a discrete-time stationary and ergodic stochastic process $X$, asymptotic equipartition property (proven for finite-valued stationary ergodic sources in Shannon-McMillan-Breiman theorem) asserts that:
\begin{equation}
	\label{eq:AEP}
    -\frac{1}{n} \log p(X_1^n) \to H(X) \quad \mbox{ as } \quad n\to\infty
\end{equation}
where $p(X_1^n)$ denotes the probability of process $X_1^n$ limited to duration $\{1,\dots, n\}$, and $H(X)$ is the entropy rate of $X$, and is shown to exist for all discrete-time stationary processes. The convergence is proven almost surely in all cases \cite{Cover:1991}. From this we can estimate $H$ by means of estimating probability of a long realisation of $X$.

Context Tree Weighting (CTW) is a data compression algorithm \cite{Willems:1995,Willems:1996,Willems:1998}, which can be interpreted as a Bayesian procedure designed to estimate the probability of a string generated by a binary tree process \cite{Gao:2008}. A binary tree process of depth $D$ is a binary stochastic process $X$ with a distribution defined with a suffix set $S$ consisting of binary strings of length $ \leq D$ and a parameter vector $\Theta = (\Theta_s ; s \in S)$, where each $\Theta_s \in [0;1]$.

If a certain string $x_1^n$ has been generated by a tree process of depth $\leq D$, but with unknown suffix set $S^*$ and parameter vector $\Theta^*$, then we may assign a prior probability $\pi(S)$ on each suffix set $S$ of depth $\leq D$ and, given $S$, we may assign a prior probability $\pi(\Theta | S)$ on each parameter vector $\Theta$. A Bayesian approximation to the true probability of $x_1^n$ (under $S^*$ and $\Theta^*$) is the mixture probability:

\begin{equation}
	\label{eq:Mixture_prob}
\hat{P}_{D, mix}(x_1^n)=\sum_s{\pi(S)}\int{P_{S,\Theta}(x_1^n)\pi(\Theta | S) d\Theta}
\end{equation}

where $P_{S,\Theta}(x_1^n)$ is the probability of $(x_1^n)$ under the distribution of a tree process with suffix set $S$ and parameter vector $\Theta$. We need an approximation since we do not know $\Theta$, and thus we cannot know the true probability. The expression in \eqref{eq:Mixture_prob} is impossible to compute directly, since the number of suffix sets of depth $\leq D$ is of order $2^D$. This is prohibitively large for practical use for any $D > 20$.

The CTW algorithm is an efficient way of computing the mixture probability in \eqref{eq:Mixture_prob} given a specific choice of the prior distributions $\pi(S), \pi(\Theta | S)$. The prior on $S$ is
\begin{equation}
	\label{eq:Prior}
\pi(S)=2^{-|S|-N(S)+1}
\end{equation}
where $|S|$ is the number of elements of $S$ and $N(S)$ is the number of strings in $S$ with length strictly smaller than $D$. Given a suffix set $S$, the prior on $\Theta$ is the product $(\frac{1}{2},\frac{1}{2})$-Dirichlet distribution, i.e., under $\pi(\Theta | S)$ the individual $\Theta_s$ are independent, with each $\Theta_s \sim$ Dirichlet$(\frac{1}{2},\frac{1}{2})$. The latter prior is conjugate in this case, while the former prior is conjugate in limit.

It's important that the CTW algorithm is able to compute the probability in \eqref{eq:Mixture_prob} precisely. The computational complexity grows linearly with the length of the string $n$. Therefore it is possible to study lengths of $D$ much higher than it is possible using plug-in estimators.

Finally, given a binary string $x_1^n$, CTW entropy rate estimator $\hat{H_{ctw}}$ is given:
\begin{equation}
	\label{eq:Entropy_CTW}
\hat{H_{ctw}}=-\frac{1}{n}\log{\hat{P}_{D,mix}(x_1^n)}
\end{equation}
where $\hat{P}_{D,mix}(x_1^n)$ is the mixture probability in \eqref{eq:Mixture_prob} computed by the CTW algorithm \cite{London:2002,Kennel:2002}. $\hat{H_{ctw}}$ will take values between $0$ for fully predictable time series and around $1$ for fully random (unpredictable) time series with alphabet of cardinality $2$, and between $0$ (predictable) and $2$ (unpredictable) for time series with alphabet of cardinality $4$.

\section{Empirical study}

In our study we estimate the entropy of the daily price time series of 358 securities traded on Warsaw Stock Exchange (GPW) (all securities which were listed for over 1000 consecutive days). The data has been downloaded from \url{http://bossa.pl/notowania/metastock/} and was up to date as of the 5th of July 2013. The data is transformed in the standard way for analysing price movements, that is so that the data points are the log ratios between consecutive daily closing prices: $r_{t}=ln(p_{t}/p_{t-1})$ and those data points are, for the purpose of the estimators, discretised into two and four distinct states. The states represent halves or quartiles, therefore each state is assigned the same number of data points. This design means that the model has no unnecessary parameters, which could affect the results and conclusions reached while using the data. This and similar experimental setups have been used in similar studies \cite{Navet:2008}. The binary alphabet ignores volatility clustering, therefore we are able to notice what part of the results stems from just this mechanism by comparing the results for binary and quadrary alphabets.

The volatility homogenisation has been designed for use with intraday returns, therefore we will also find the predictability of high frequency price changes estimated for another set of data, that is intraday price changes (transaction level) for 707 securities listed on Warsaw Stock Exchange which had over 2500 price changes recorded (all securities with at least 2500 recorded price changes). Unlike in the daily data we have ignored data points where there was no price change, as those are not meaningful for data points separated by very small (not uniform) time intervals. Thus we estimate predictability of the next price change. On the daily data such omission would be irrelevant.

On the log returns themselves we have performed the volatility homogenisation decomposition for various values of $\delta$: $0.05$, $0.1$, $0.25$, $0.5$, $0.75$ \& $1$. It's obvious that price shifts of one whole monetary unit, which we are looking at with the last mentioned value of $\delta$, are happening at a very low rate, thus the resulting time series will be much shorter than the original ones. We are therefore limiting our results for the intraday data to 265 times series with at least 1000 changes for $\delta=1$. It is motivated by the needs of the CTW estimator, which will not give credible results for very short time series.

\begin{figure}[tbh]
\centering
\includegraphics[width=0.35\textwidth]{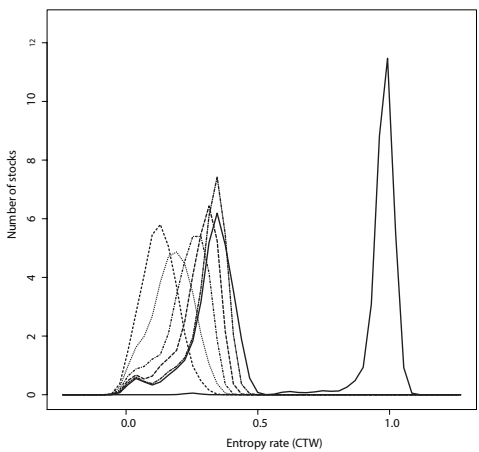}
\caption{Kernel densities of Shannon's entropy rates estimated for intraday log returns, particularly from left to right: volatility homogenisation decomposed binary time series with $\delta$: $0.05$, $0.1$, $0.25$, $0.5$, $0.75$ \& $1$ and original binary time series. High values of entropy rates denote low predictability.}
\label{fig:SM1}
\end{figure}

\begin{figure}[tbh]
\centering
\includegraphics[width=0.35\textwidth]{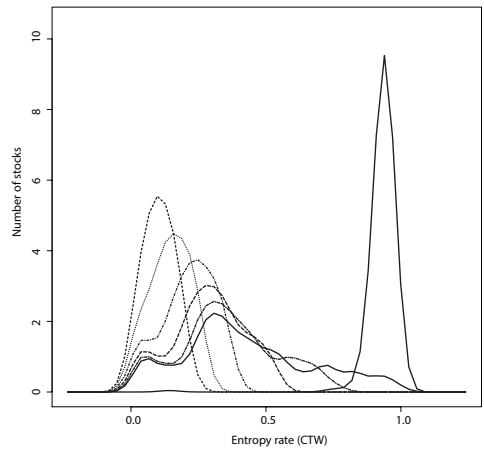}
\caption{Kernel densities of Shannon's entropy rates estimated for daily log returns, particularly from left to right: volatility homogenisation decomposed binary time series with $\delta$: $0.05$, $0.1$, $0.25$, $0.5$, $0.75$ \& $1$ and original binary time series. High values of entropy rates denote low predictability.}
\label{fig:SM2}
\end{figure}

First, we want to see whether the decomposed time series are more or less predictable than the original ones. For this purpose we compare the distributions (in the form of kernel densities) of the Shannon's entropy rates for the 265 stocks for the original times series discretised into a binary alphabet (we are not comparing with an alphabet of four letters as then the entropy rate would have a different scale, since volatility homogenisation gives us binary time series), and the decomposed series for the above mentioned values of $\delta$. In Fig.~\ref{fig:SM1} we present the kernel densities for the intraday time series. From left to right these represent volatility homogenisation decomposed binary time series with $\delta$: $0.05$, $0.1$, $0.25$, $0.5$, $0.75$ \& $1$ and the last one on the right represents the original time series discretised into two states. In Fig.~\ref{fig:SM2} we present the kernel densities for the daily time series (358 securities). Similarly, from left to right these represent volatility homogenisation decomposed binary time series with $\delta$: $0.05$, $0.1$, $0.25$, $0.5$, $0.75$ \& $1$ and the last one on the right represents the original time series discretised into two states. Once again we note that low values of entropy rates denote high predictability.

\begin{figure}[tbh]
\centering
\includegraphics[width=0.35\textwidth]{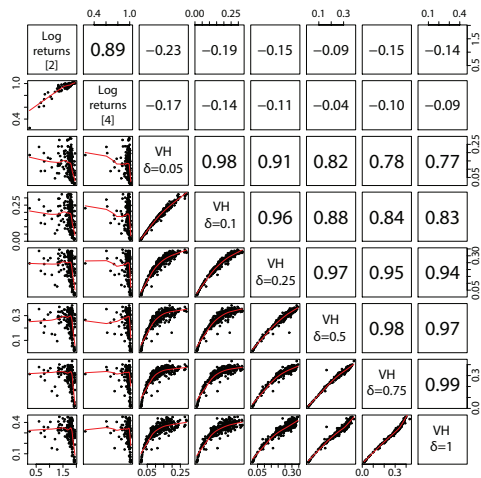}
\caption{Pearson's correlation between Shannon's entropy rates estimated for intraday log returns: original binary time series, original quadrary time series, and volatility homogenisation decomposed binary time series with $\delta$: $0.05$, $0.1$, $0.25$, $0.5$, $0.75$ \& $1$.}
\label{fig:Co1}
\end{figure}

\begin{figure}[tbh]
\centering
\includegraphics[width=0.35\textwidth]{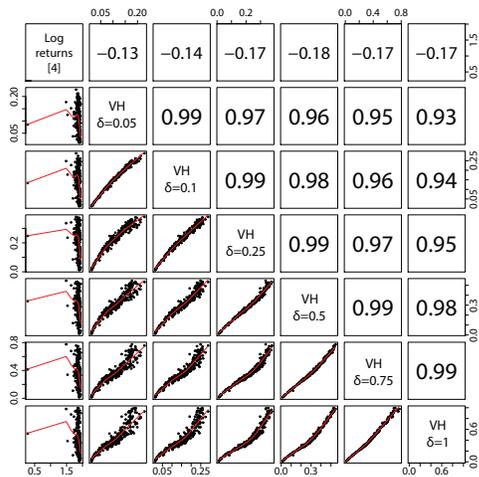}
\caption{Pearson's correlation between Shannon's entropy rates estimated for daily log returns: original binary time series, original quadrary time series, and volatility homogenisation decomposed binary time series with $\delta$: $0.05$, $0.1$, $0.25$, $0.5$, $0.75$ \& $1$.}
\label{fig:Co2}
\end{figure}

Second, we want to see what is the robustness of this method with regards to the choice of the parameters, or in other words we want to see whether the choice of the parameter $\delta$ changes the results dramatically or just finely tunes them. For this purpose, in Fig.~\ref{fig:Co1} we show the Pearson's correlation coefficients between entropy rates calculated for the original intraday time series discretised into two and four states and for the decomposed intraday time series with the above mentioned values of $\delta$. In Fig.~\ref{fig:Co2} we show the same for daily time series.

\begin{figure}[tbh]
\centering
\includegraphics[width=0.35\textwidth]{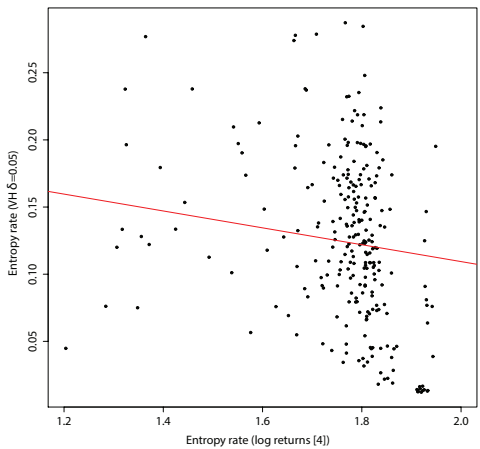}
\caption{Scatterplot of Shannon's entropy rates estimated for intraday log returns based on original quadrary time series and volatility homogenisation decomposed time series with $\delta=0.05$.}
\label{fig:Sc1}
\end{figure}

\begin{figure}[tbh]
\centering
\includegraphics[width=0.35\textwidth]{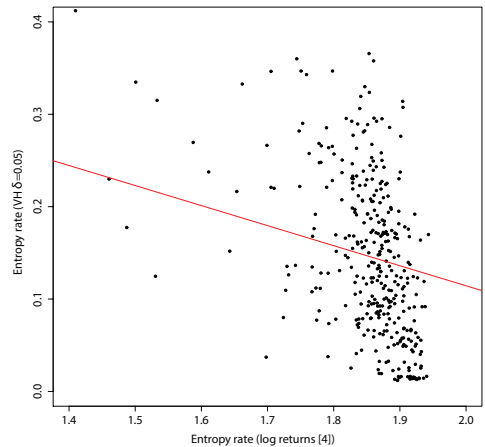}
\caption{Scatterplot of Shannon's entropy rates estimated for daily log returns based on original quadrary time series and volatility homogenisation decomposed time series with $\delta=0.05$.}
\label{fig:Sc2}
\end{figure}

Lastly, we show visually (using a scatterplot) how the negative correlation between Shannon's entropy rates for original and decomposed time series presented above is realised, as to get a more intuitive understanding of what it means. In Fig.~\ref{fig:Sc1} we present the relationship between Shannon's entropy rates calculated for original daily time series discretised into four states and volatility homogenisation decomposed time series with $\delta=0.05$. In Fig.~\ref{fig:Sc2} we present the same value for daily time series.

%NYSE???

\section{Discussion}

In Fig.~\ref{fig:SM1} we see that while the time series describing log returns are highly unpredictable, hovering around them maximum entropy of $1$, the decomposed time series are much more predictable, and in fact this predictability is higher for low values of $\delta$. The average Shannon's entropy rate for various values of $\delta$ are presented in Table~\ref{table.1}. This is desirable, as with high values of $\delta$ the even times will be far apart in real market times, thus rendering such time series less interesting for practical purposes. It is also what we would expect as with small values of $\delta$ there will be a significant number of price changes much larger than $\delta$, which will constitute patterns of several $\delta$ sized movements in the same direction in the decomposed data, that is patterns in themselves. In other words it is harder to predict large price shifts than it it to predict small price movements. We are therefore inclined to say that the volatility homogenisation decomposition does render the time series much less noisy, leaving us with data filled with patterns, which, in principle, can be used for profit. Judging by the mentioned results in Ref. \cite{Kowalewski:2014} this may not be always an easy task, which is confirmed by the studies looking into the relationship between predictability and profitability \cite{Fiedor:2014}, nonetheless it is possible in principle. In Fig.~\ref{fig:SM2} we can see that a similar pattern emerges for daily stock returns on the Warsaw's market. This time the results for the decomposed time series are much less stable however, which is understandable given that this method has been designed to deal with high frequency data. It does appear that for low values of $\delta$ the method may be used for daily returns as well, even though the usefulness of such an approach is an open question.

\begin{table}[htbp]
\caption{Average Shannon's entropy rate for all studied time series for various values of parameter $\delta$}
\begin{ruledtabular}
\begin{tabular}{lr}
%\hline
\multicolumn{1}{l}{$\delta$} & \multicolumn{1}{l}{$\bar{H_{ctw}}$} \\ \hline
0.05 & 0.13 \\
0.10 & 0.17 \\
0.25 & 0.23 \\
0.50 & 0.27 \\
0.75 & 0.30 \\
1.00 & 0.32 \\
\end{tabular}
\end{ruledtabular}
\label{table.1}
\end{table}

In Fig.~\ref{fig:Co1} we see that the entropy rates estimated for decompositions with various values of $\delta$ are highly correlated, leading us to believe that volatility homogenisation is not sensitively dependent on the choice of this parameter. It is important since in practical applications the choice of this parameter does not stem from any theory and is largely a question of the intuition of the analyst and the particular interest in a given price change granularity. In Fig.~\ref{fig:Co2} we see that similar results appear for daily returns, and in fact here the correlation is much stronger, ranging from $0.93$ to $0.99$, whereas the Pearson's correlation coefficient between Shannon's entropy rates estimated for the time series decomposed with $\delta$ equal $0.05$ and $1$ is equal to $0.77$. Thus in principle an analyst can choose any value of $\delta$ they are interested in from their perspective and use this method with good results. The sensitivity of the used predictive models to these small changes may exist for certain methods however, but this problem is outside the scope of this paper. On these figures it is particularly interesting to see that it appears the less entropic original time series lead to slightly more entropic decomposed time series, and vice versa. This could be helpful, since the time series with very few patterns could be turned into times series with very many patterns. To find out why there is such a correlation coefficient present we plot the relationship between Shannon's entropy rates estimated for original time series discretised into quartiles and decomposed time series with $\delta=0.05$ for intraday and daily returns in Figs.~\ref{fig:Sc1} \& \ref{fig:Sc2} respectively. We can see that this appears to be caused mostly by the much larger standard deviation of the Shannon's entropy rates for the decomposed time series. The original time series appear to be more coherent in their unpredictability. It appears that while the predictability of the original time series is mostly defined by the noise in the data (ie. they are almost completely unpredictable), the predictability of the decomposed time series is much more dependent on the underlying structure of the returns rather than the noise, making the bigger change in predictability between original and decomposed time series for stocks with the most noise in the original log returns, as well making the predictability of decomposed time series much more varied among the stocks.

In closing we will also note that we have performed the same study on time series describing NYSE100 stocks on daily time scale as of 11th of November 2013, going 10 years back, and on one minute scale, for 15 days between the 21st October 2013 and the 8th of November 2013. These time series are much shorter, and the number of stocks is smaller as well. This makes the results less convincing and less stable, thus these results were not reported, especially since they would agree with our reported findings. In fact the results were virtually identical to the first approximation in all the above considerations. We are therefore postulating that these results are not specific to the Warsaw's market but are indeed inherent to all financial markets.

\section{Conclusions}

In this study we have shown that volatility homogenisation decomposition does indeed greatly enhance predictability of the time series describing stock returns. In the case of intraday data, for which the method has been designed, the choice of the width of the studied regular points in space only slightly alters the result, thus we have also shown that the volatility homogenisation is also robust with regards to the choice of this parameter. This is futher corroborated by the correlation analysis presented. Additionally it seems that the more unpredictable the original stock returns are the more predictable the volatility homogenisation decomposition will be, perhaps hinting that the noise in the original log returns time series is well removed by this method. Our analysis was performed on a very large database of intraday and daily stock returns on Warsaw Stock Exchange. We postulate, based on our previous research as well as a limited replication of this study on NYSE100 data, that these should be representative for other markets as well. Nonetheless, further research could attempt to replicate these results for New York Stock Exchange and other European markets. Further studies should then try to apply this decomposition with various time series to compare its usefulness with different methods of prediction, since, as has been shown \cite{Fiedor:2014}, the connection between predictability and profitability is not fully understood. Further research should also be performed to assess the usefulness of such decomposition of time series in other fields where it may be beneficial.

\bibliography{prace}
\end{document}